\documentclass[iop]{emulateapj}
\usepackage{amsmath}
\usepackage{amssymb}
\usepackage{bm}
\usepackage{graphicx}
\usepackage{natbib}

\newcommand{\R}{\mathbb{R}}
\newcommand{\Z}{\mathbb{Z}}
\renewcommand{\mod}{\;\mathrm{mod}\;}
\renewcommand{\vec}[1]{\bm{#1}}
\newcommand{\pmat}[1]{\begin{pmatrix} #1 \end{pmatrix}}

\begin{document}

\shorttitle{Remapping simulations}
\shortauthors{Carlson \& White}
\title{Embedding realistic surveys in simulations through volume remapping}
\author{Jordan Carlson \& Martin White}
\affil{Department of Physics, University of California, Berkeley, CA 94720}

\date{\today}

\begin{abstract}
Connecting cosmological simulations to real-world observational programs is
often complicated by a mismatch in geometry: while surveys often cover highly
irregular cosmological volumes, simulations are customarily performed in a
periodic cube.  We describe a technique to remap this cube into elongated
box-like shapes that are more useful for many applications.  The remappings are
one-to-one, volume-preserving, keep local structures intact, and involve
minimal computational overhead.
\end{abstract}

\keywords{methods: $N$-body simulations ---
          cosmology: large-scale structure of universe}

\section{Introduction}

Numerical simulations have become an indispensable tool in modern cosmological
research, used for investigating the interplay of complex physical processes,
studying regimes of a theory which cannot be attacked analytically, generating
high precision predictions for cosmological models, and making mock catalogs
for the interpretation and analysis of observations.  Such simulations
traditionally evolve the matter distribution in periodic cubical volumes, which
neatly allow them to approach the homogeneous Friedmann solution on large
scales.  The use of a periodic volume also allows the long-range force to be
easily computed by fast Fourier transform methods in many popular algorithms
(e.g.~particle-mesh, particle-particle-particle-mesh or tree-particle-mesh
algorithms).

Surveys, on the other hand, often cover cosmological volumes that are far from
cubical in shape, and making a mock catalog which includes the full geometrical
constraints of the observations is difficult.  One approach is to simulate a
sufficiently large volume that the survey can be embedded directly within the
cube, but this often means large parts of the computational domain are unused.
An alternative is to trace through the cube across periodic boundaries so as to
generate the desired depth, with various rules for avoiding replication or
double-counting of the volume (if desired).  A third approach is to run a
simulation in a non-cubical geometry.  If the side lengths are highly
disproportionate this can lead to its own numerical issues, and in addition it
makes it difficult to reuse a given simulation for many applications.

\begin{figure}
\begin{center}%
\resizebox{7cm}{!}{\includegraphics{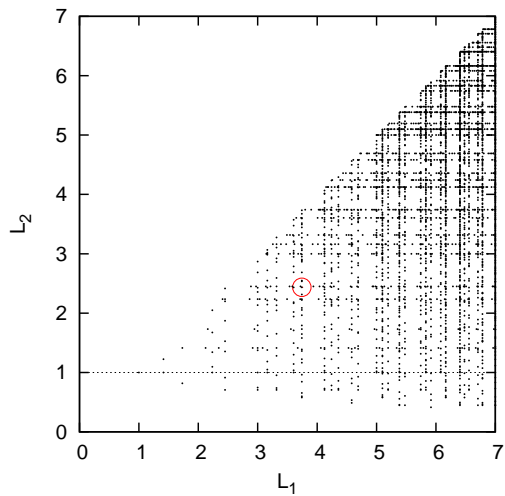}}%
\end{center}%
\vspace{-0.15in}
\caption{Possible dimensions $L_1 \times L_2 \times L_3$ for cuboid remappings,
subject to the conditions $L_1\ge L_2\ge L_3$, $L_1 L_2 L_3=1$, and $L_1<7$.
The choice used for the Stripe 82 remapping described in Section
\protect\ref{sec:examples} is indicated by a red circle.}
\label{fig:poss}
\end{figure}

In this paper we present a new solution to this problem which allows one to
embed a hypothetical survey volume inside a cosmological simulation while
limiting wasted volume and artificial correlations.  The method is based on the
simple observations that
(1) a cube with periodic boundary conditions is equivalent to an infinite
3-dimensional space with discrete translational symmetry, and
(2) the primitive cell for such a space need not be a cube.
We show that one may take the primitive cell to be a cuboid%
\footnote{To be precise, by \emph{cuboid\/} we mean here a rectangular
parallelepiped, i.e.~a parallelepiped whose six faces are all rectangles
meeting at right angles.  For practical purposes, a cuboid is simply a box that
is not a cube.}
of dimensions $L_1 \times L_2 \times L_3$ for a discrete but large choice of
values $L_1\ge L_2\ge L_3$.  The possible choices, subject to the constraint
$L_1 L_2 L_3 = 1$, are illustrated in Figure \ref{fig:poss}.

Our approach leads to a one-to-one remapping of the periodic cube which keeps
structures intact, does not map originally distant pieces of the survey close
together, and uses no piece of the volume more than once.  It complements
existing techniques for generating mock observations (e.g.~%
\citealt{Blaizot+05,Kitzbichler+07}) or for ray-tracing through simulations
(e.g.~\citealt{WhiHu00,Vale+03,Hilbert+09,Fullana+10}).  Though not ideal in
all cases, our remapping procedure may be seen as a general-purpose alternative
that neatly skirts many of the complications involved with previous methods.

We begin in Section \ref{sec:math} with a mathematical description of the
remapping, and explain what choices of dimensions are possible.  In Section
\ref{sec:algorithm} we describe how to implement this remapping numerically.
We present a few useful examples in Section \ref{sec:examples}, and conclude in
Section \ref{sec:discussion} with a discussion of some of the advantages and
limitations of this method.

\section{Mathematical description} \label{sec:math}

Consider a unit cube with periodic boundary conditions.  By tiling copies of
this cube in all directions, each point $(x,y,z) \in \R^3$ corresponds to a
point in the canonical unit cube $[0,1]^3$ via the map
\begin{equation}
    (x,y,z) \mapsto (x \mod{1},~ y \mod{1},~ z \mod{1}),
    \label{eq:mod}
\end{equation}
where ``$x \mod{1}$'' is just the fractional part of $x$ if $x \ge 0$, or one
minus this fractional part for $x < 0$.  Since we can identify each point in
$\R^3$ with a point in $[0,1]^3$, any region in space corresponds to a sampling
of the unit cube.  We are primarily interested in bijective samplings, i.e.\
regions that according to the above equivalence cover each point of the unit
cube once and only once.  We call such a sampling a \emph{remapping\/} of the
unit cube.  Equivalently, a remapping may be thought of as a partitioning of
the unit cube into disjoint regions, which are then translated by integer
offsets and glued back together along periodic boundaries.

\begin{figure*}
\begin{center}
\includegraphics{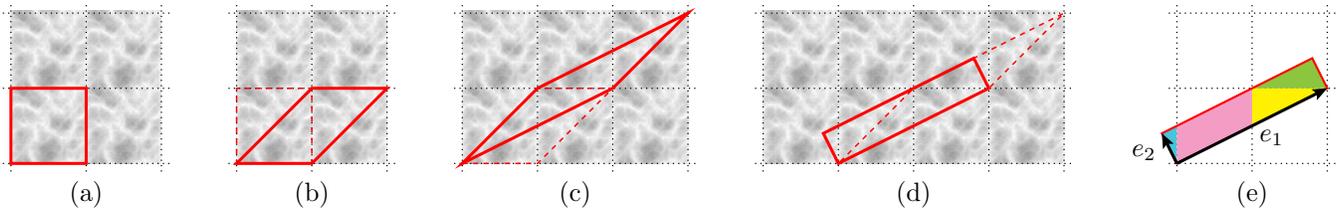}
\end{center}
\caption{Panels (a)-(d) show a sequence of shear transformations that results
in a remapping of the unit square into a rectangle of dimensions $\sqrt{5}
\times 1/\sqrt{5}$, as described in the text.  Panel (e) highlights the edge
vectors $\vec{e}_1, \vec{e}_2$, and the four cells that comprise this
rectangle.  Note that each point of the original unit square is covered exactly
once in the final remapping.}
\label{fig:steps}
\end{figure*}

A general class of such remappings may be constructed by applying shear
transformations to the unit cube.  We first illustrate the idea in 2D, where
the transformations may be easily visualized, and then state the appropriate
generalization to 3D.  Start with a continuous field laid down within the unit
square, which may be thought of as a parallelogram spanned by the vectors
$\vec{u}_1 = (1,0)$ and $\vec{u}_2 = (0,1)$.  We suppose the field to be
periodic, and tile copies throughout the plane [see Figure \ref{fig:steps}(a)].
Now imagine taking the unit square and shearing it along its top edge, leaving
the underlying field in place [Figure \ref{fig:steps}(b)].  The result is a new
parallelogram defined by the vectors
\begin{equation}
    \vec{u}_1 = (1,0), \qquad \vec{u}_2' = \vec{u}_2 + m \vec{u}_1 = (m,1),
\end{equation}
where $m$ is some real scalar that controls the extent of the shear.  Since shear
transformations are area-preserving, this parallelogram has unit area.  In
fact, since $\vec{u}_1$ is a lattice vector (i.e.~a vector of translational
invariance), this parallelogram covers each point of our field once and
only once, and hence defines a valid remapping.

Now observe that if $m$ is an integer, then $\vec{u}_2' = (m,1)$ will also be a
lattice vector.  In that case we again have a parallelogram with edge vectors
$\vec{u}_1$ and $\vec{u}_2'$ that are both lattice vectors.  We can now shear
this parallelogram along its right edge [Figure \ref{fig:steps}(c)], giving a
new parallelogram with edge vectors
\begin{equation}
    \vec{u}_1' = \vec{u}_1 + n \vec{u}_2' = (1+mn,n), \qquad \vec{u}_2' = (m,1).
\end{equation}
Once again this parallelogram covers each point of our field exactly once, and
if $n$ is an integer then $\vec{u}_1'$ and $\vec{u}_2'$ are both lattice
vectors.  We may repeat this process, applying integer shears alternately to
the top and right edges of the parallelogram.  In general, if $\vec{u}_1$ and
$\vec{u}_2$ are \emph{any\/} integer-valued 2D vectors that span a
parallelogram of unit area, they can be obtained by such a sequence of shear
transformations.  This condition is simply
\begin{equation}
    u_{11} u_{22} - u_{12} u_{21} =
    \det \pmat{u_{11} & u_{12} \\ u_{21} & u_{22}} = 1,
\end{equation}
i.e.~$\vec{u}_1$ and $\vec{u}_2$ are the rows of an invertible integer-valued
matrix.

While remapping the unit square into a parallelogram may be useful in certain
cases, parallelograms are generally too awkward to be useful.  Instead, given
lattice vectors $\vec{u}_1$ and $\vec{u}_2$ spanning unit area, we may apply
one last shear to ``square up'' the parallelogram into a rectangle [Figure
\ref{fig:steps}(d)].  Explicitly, we let
\begin{equation}
    \vec{e}_1 = \vec{u}_1, \qquad \vec{e}_2 = \vec{u}_2 + \alpha \vec{u}_1,
\end{equation}
and choose $\alpha$ so that $\vec{e}_1$ and $\vec{e}_2$ are orthogonal.  The
rectangle defined by $\vec{e}_1$ and $\vec{e}_2$ now covers each point of the
unit square exactly once.

The generalization to 3D is straightforward.  By applying integer shears to the
faces of the cube, we may obtain any parallelepiped with edges given by integer
vectors $\vec{u}_1$, $\vec{u}_2$, $\vec{u}_3$ satisfying
\begin{equation}
   \det \pmat{u_{11} & u_{12} & u_{13} \\ u_{21} & u_{22} & u_{23} \\ u_{31} & u_{32} & u_{33}} = 1.
\end{equation}
Again this space of possibilities corresponds to the space of invertible,
integer-valued $3\times3$ matrices.  We apply two final shears to square up
this parallelepiped into a cuboid, by choosing coefficients $\alpha$, $\beta$,
and $\gamma$ such that
\begin{align}
    \vec{e}_1 &= \vec{u}_1, \\
    \vec{e}_2 &= \vec{u}_2 + \alpha \vec{u}_1, \\
    \vec{e}_3 &= \vec{u}_3 + \beta \vec{u}_1 + \gamma \vec{u}_2,
\end{align}
are mutually orthogonal.  This gives a remapping of the unit cube into a cuboid
with side lengths $L_i = |\vec{e}_i|$, satisfying $L_1 L_2 L_3 = 1$.  Moreover
since $\vec{e}_1$ is still a lattice vector, this cuboid is periodic across the
face perpendicular to this vector.

Invertible, integer-valued $3\times3$ matrices can be generated easily by brute
force computation, and each such matrix leads to a cuboid remapping of the unit
cube.  The allowable dimensions $L_1\times L_2\times L_3$ for these cuboids are
illustrated in Figure \ref{fig:poss}, where $L_3$ is given implicitly by the
unit volume condition $L_1 L_2 L_3 = 1$.

\section{Numerical algorithm} \label{sec:algorithm}

The goal of our remapping procedure is to provide an explicit bijective map
between the unit cube and a cuboid of dimensions $L_1\times L_2\times L_3$.  We
will refer to points in the unit cube by their \emph{simulation coordinates\/}
$\vec{x} \in [0,1]^3$, and their remapped positions in a canonical,
axis-aligned cuboid by \emph{remapped coordinates\/} $\vec{r} \in
[0,L_1]\times[0,L_2]\times[0,L_3]$.

The reverse map $\vec{r} \mapsto \vec{x}$ is simple.  The edge vectors
$\vec{e}_1, \vec{e}_2, \vec{e}_3$ discussed previously describe how to embed an
oriented cuboid within an infinite tiling of the unit cube.  Let
$\hat{n}_i = \vec{e}_i/L_i$ be unit vectors along these edges.  Then given
$(r_1,r_2,r_3) \in [0,L_1]\times[0,L_2]\times[0,L_3]$, the point
\begin{equation}
    \vec{p} = r_1 \hat{n}_1 + r_2 \hat{n}_2 + r_3 \hat{n}_3
\end{equation}
lies within this oriented cuboid, and this maps to a point in the unit cube
according to Eq. (\ref{eq:mod}).

The forward map $\vec{x} \mapsto \vec{r}$ is slightly more complicated.
Consider again the oriented cuboid with edge vectors $\vec{e}_1, \vec{e}_2,
\vec{e}_3$ embedded within an infinite tiling of the unit cube.  Each tile
(i.e.~each replication of the unit cube) may be labeled naturally by an integer
triplet $\vec{m} \in \Z^3$, where the canonical unit cube has
$\vec{m} = (0,0,0)$.  The intersection of the cuboid with a tile is called a
\emph{cell\/}, of which only a finite number will be non-empty.  Each non-empty
cell is uniquely labeled by the triplet $\vec{m}$.  [The four cells of the 2D
example from the previous section are indicated in Figure \ref{fig:steps}(e).]

As a geometrical object, each cell is just a convex polyhedron bounded by 12
planes: the 6 faces of the tile and the 6 faces of the cuboid.  A plane may be
parametrized by real numbers $(a,b,c,d)$, so that a point $(x,y,z)$ lies
inside, outside, or on the plane depending on whether the quantity $ax + by +
cz + d$ is less than, greater than, or equal to zero.  Thus to test whether a
point belongs to a cell, we need only check if it lies inside all the planes
that bound it.  When translated spatially by a displacement $-\vec{m}$, each
cell defines a region within the unit cube.  The collection of all such cells
defines a partitioning of the unit cube, with each point in $[0,1]^3$ being
covered by exactly one cell.  Our algorithm for the forward map $\vec{x}
\mapsto \vec{r}$ then may be summarized as follows:
\begin{enumerate}
\item determine which cell contains the point $\vec{x}$ (using point-plane tests)
\item let $\vec{p} = \vec{x} + \vec{m}$ be the corresponding point in the oriented cuboid
\item define $\vec{r} \in [0,L_1]\times[0,L_2]\times[0,L_3]$ by $r_i = \vec{p}\cdot\hat{n}_i$
\end{enumerate}

\section{Examples and applications} \label{sec:examples}

The mappings described above can be used in myriad ways.  The algorithm is fast
enough that it can be used for on-the-fly analysis while a simulation is
running, it introduces little overhead in walking a merger tree (allowing
complex light-cone outputs to be built during e.g.~the running of a
semi-analytic galaxy formation code), or it can be applied in post-processing
to a static time output of a simulation to alter the geometry.  In this section
we give a (very) few illustrative examples.

\begin{figure}
\begin{center}
\resizebox{2.8in}{!}{\includegraphics{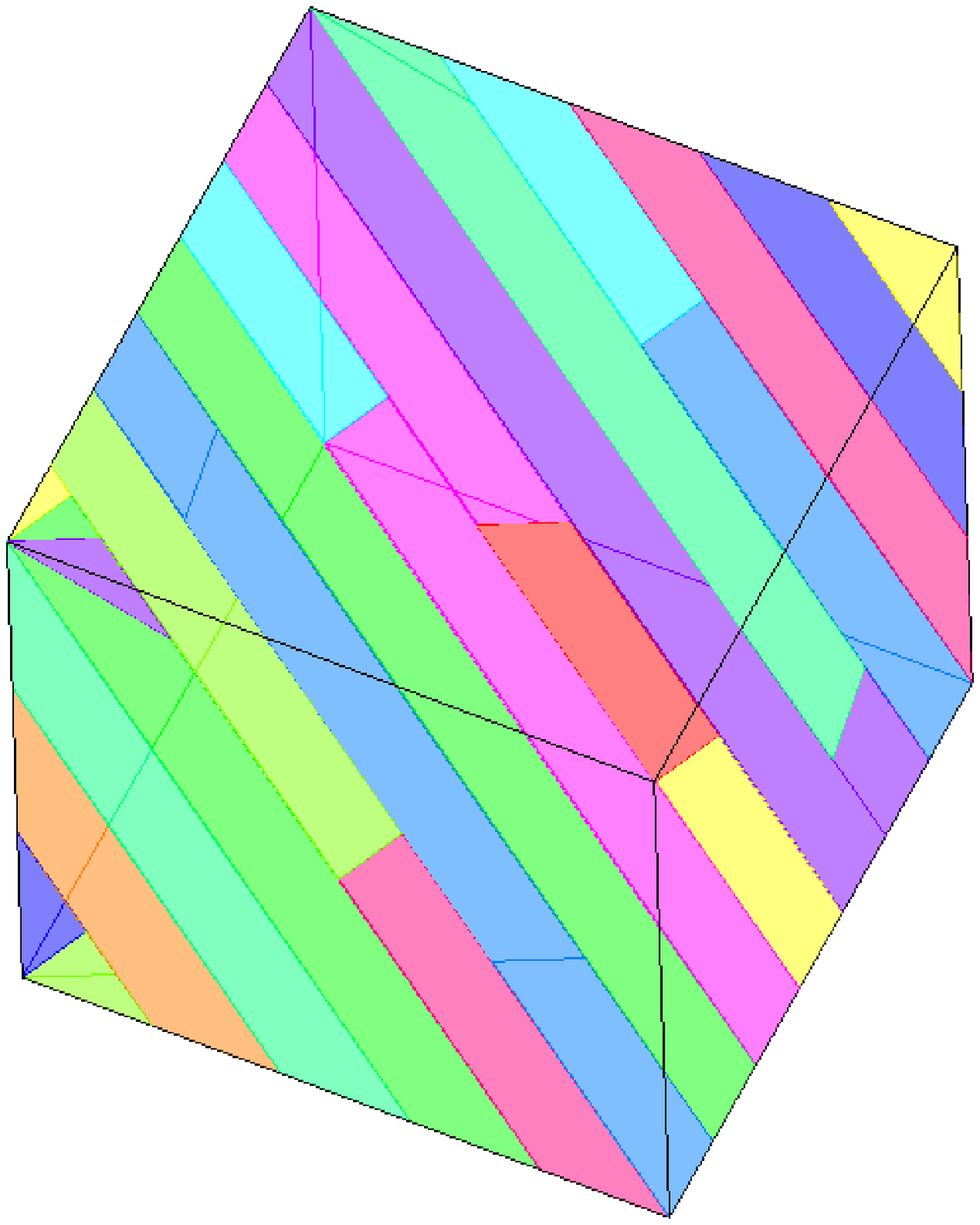}}
\resizebox{2.8in}{!}{\includegraphics{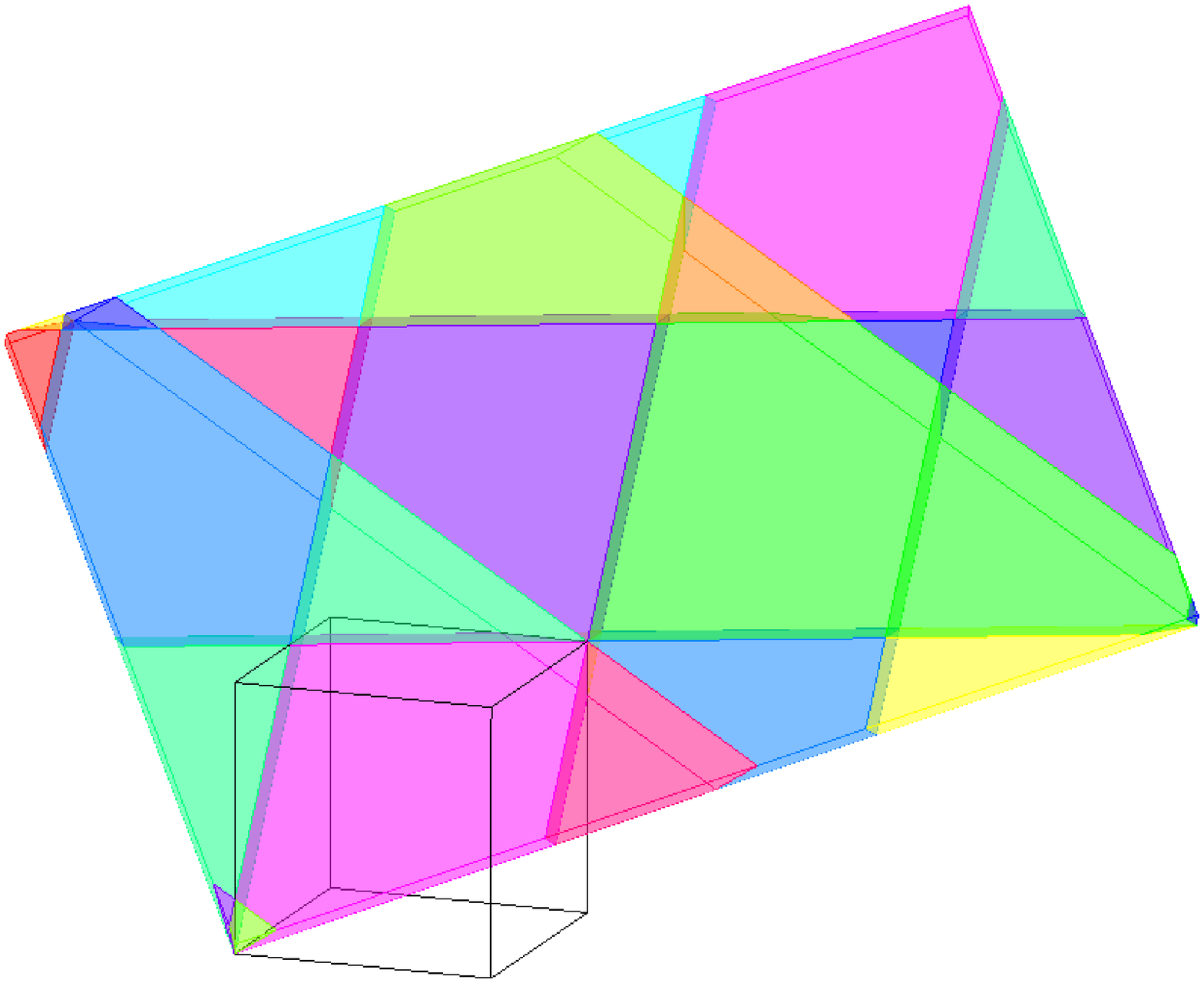}}
\end{center}
\caption{The regions in the unit cube, and how they ``unfold'' to produce
the wide, thin geometry shown in Figure \protect\ref{fig:stripe82}.}
\label{fig:3dview}
\end{figure}

As it is so widely known we shall use as our fiducial volume the Millennium
simulation \citep{MS}, which was performed in a cubical volume of side length
$500\,h^{-1}$Mpc.  We begin by asking how we could embed a very wide-angle
survey, such as the equatorial stripe (``Stripe 82'') of the Sloan Digital Sky
Survey\footnote{{\tt http://www.sdss.org}}, in such a volume.  Stripe 82 is
$100^\circ$ wide and $2.5^\circ$ in height.  Using the Millennium simulation
cosmology, the volume within Stripe 82 out to $z\simeq 0.45$ equals the total
volume within the Millennium simulation.  At this depth the stripe is
$1.2\,h^{-1}$Gpc in the line-of-sight direction, $1.9\,h^{-1}$Gpc transverse
but only $50\,h^{-1}$Mpc deep.

\begin{figure*}
\begin{center}
\resizebox{12cm}{!}{\includegraphics{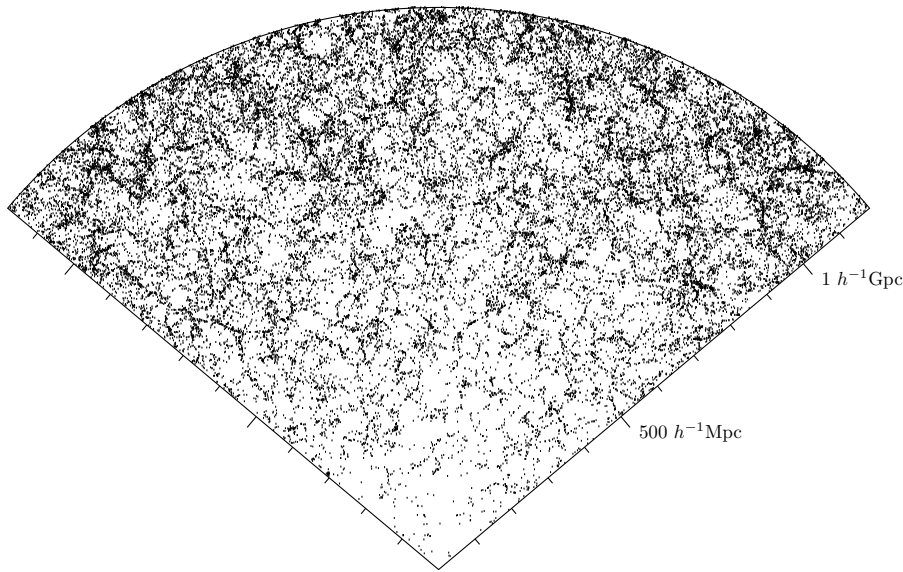}}
\end{center}
\caption{A mock catalog from the Millennium simulation remapped into the
equatorial stripe of the Sloan Digital Sky Survey.  The ``stripe'' is a thin
rectangle on the sky of angular dimensions $100^\circ\times2.5^\circ$, shown
here ``from above'' extending out to redshift $z\simeq 0.45$, or $r\simeq
1.2\,h^{-1}$Gpc comoving.  We show only galaxies brighter than $R=-20.5$, to
avoid saturating the figure.}
\label{fig:stripe82}
\end{figure*}

We can map the cubical volume of the simulation into this geometry using the
transformation circled in red in Fig.~\ref{fig:poss}.  A view of this
transformation is given in Figure \ref{fig:3dview}, where the regions within
the cube are shown ``unfolded'' to produce a long and wide, but thin, domain
into which we can embed the Stripe 82 geometry.  Figure \ref{fig:stripe82}
shows a light-cone produced from the outputs of the Millennium simulation, in
the Stripe 82 geometry.  Specifically we show all galaxies brighter than
$R=-20.5$ (to avoid saturating the figure) and with $0<z<0.45$ from the catalog
of \citet{deLucia+07}.

Another frequently encountered situation is a survey which is much longer in
the line-of-sight direction than either of the (approximately equal) transverse
directions.  Let us consider for example a survey which aims to reach $z\simeq
3$, or about $5\,h^{-1}$Gpc in the Millennium simulation cosmology.  Among the
many possible choices at our disposal we find a remapping with
$(L_1,L_2,L_3) = (5025,179,139)\,h^{-1}$Mpc, which could encompass a survey of
angular dimensions $2.0^\circ\times1.6^\circ$ out to $z \simeq 3.5$.  To probe
even earlier epochs we could choose a remapping with
$(L_1,L_2,L_3) = (7106,145,121)\,h^{-1}$Mpc, which allows a
$1.2^\circ\times1.0^\circ$ survey out to $z \simeq 10$.  For these types of
highly elongated geometries, the possible remappings are dense enough that a
suitable choice may be found for almost any survey.


\section{Discussion} \label{sec:discussion}

As surveys become increasingly complex and powerful and the questions we ask of
them become increasingly sophisticated, mock catalogs and simulations which can
mimic as much as possible the observational non-idealities become increasingly
important.  \citet{Angulo+10} have shown that simulations of one cosmology can
be rescaled to approximate those of a different cosmology.  We have introduced
a remapping of periodic simulation cubes which allows one simulation to take on
the characteristics of many different observational geometries.  The use of
such techniques enhances the usefulness of cosmological simulations, which
often involve a large investment of community resources.

The methods we introduced in this paper lead to one-to-one remappings of the
periodic cube which keep structures intact, do not map originally distant
pieces of the survey close together, and use no piece of the volume more than
once.  The remapping can be done extremely quickly, meaning it can be included
in almost any analysis tool with negligible overhead.

Remapping the computational geometry is, however, not without its limitations.
First and foremost, although the target geometry may have sides much longer
than the original simulation, the structures will not contain the correct large
scale power since it was missing from the simulation to begin with.  This
problem becomes less acute as the original simulation volume becomes a fairer
representation of the Universe.  Secondly, if the target geometry is too thin
it is possible for points which are far apart in the survey to come from points
close together in the simulation volume, leading to spurious correlations.
These are analogous to the artificial correlations between survey ``sides''
that occur when it is embedded in a periodic cube.  Simply excluding a boundary
layer from the remapped volume can tame such correlations.

In addition to the description of the algorithm in this paper, we have made
Python and C++ implementations of the remappings, along with further examples
and animations, publicly available at {\tt http://mwhite.berkeley.edu/BoxRemap}.

\vspace*{1cm}

\begin{acknowledgments}
We thank the members of the DES simulation working group for valuable feedback
and encouragement.
This work was supported by NASA and the DoE.
\end{acknowledgments}

\bibliography{boxremap}{}
\bibliographystyle{apj}

\end{document}